\documentclass[prl,twocolumn]{revtex4-1}
\usepackage{graphicx}
\usepackage{amsmath}
\usepackage{amssymb}
\usepackage{textcomp}
\usepackage{color}
\usepackage{transparent}

\begin{document}
\title{Blue-detuned magneto-optical trap}
\author{K. N. \surname{Jarvis}}
\email{k.jarvis14@imperial.ac.uk}
\affiliation{Centre for Cold Matter, Blackett Laboratory, Imperial College London, Prince Consort Road, London SW7 2AZ, UK}
\author{J. A. Devlin}
\affiliation{Centre for Cold Matter, Blackett Laboratory, Imperial College London, Prince Consort Road, London SW7 2AZ, UK}
\author{T. E. Wall}
\affiliation{Centre for Cold Matter, Blackett Laboratory, Imperial College London, Prince Consort Road, London SW7 2AZ, UK}
\author{B. E. Sauer}
\affiliation{Centre for Cold Matter, Blackett Laboratory, Imperial College London, Prince Consort Road, London SW7 2AZ, UK}
\author{M. R. Tarbutt}
\affiliation{Centre for Cold Matter, Blackett Laboratory, Imperial College London, Prince Consort Road, London SW7 2AZ, UK}
\begin{abstract}
We present the properties and advantages of a new magneto-optical trap (MOT) where blue-detuned light drives `type-II' transitions that have dark ground states. Using $^{87}$Rb, we reach a radiation-pressure-limited density exceeding $10^{11}$~cm$^{-3}$ and a temperature below 30~$\mu$K. The phase-space density is higher than in normal atomic MOTs, and a million times higher than comparable red-detuned type-II MOTs, making it particularly attractive for molecular MOTs which rely on type-II transitions. The loss of atoms from the trap is dominated by ultracold collisions between Rb atoms. For typical trapping conditions, we measure a loss rate of $1.8(4)\times10^{-10}$~cm$^{3}$~s$^{-1}$. 
\end{abstract}
\maketitle

The magneto-optical trap (MOT)~\cite{Prentiss1987} is an essential tool for a wealth of scientific and technological applications of ultracold atoms, including tests of fundamental physics, studies of ultracold collisions and quantum degenerate gases, advances in frequency metrology and the development of commercial cold atom instruments such as gravimeters and clocks. Similarly, applications of ultracold molecules~\cite{Carr2009} are sure to be advanced by recent demonstrations of molecular MOTs~\cite{Barry2014, Norrgard2016, Truppe2017b, Anderegg2017}. All previous MOTs have been made using light red-detuned from the atomic or molecular transition. This is a requirement for Doppler cooling, the primary cooling mechanism in a MOT. Sub-Doppler cooling also requires red-detuned light when the excited state angular momentum ($F'$) exceeds that of the ground state ($F$). These transitions are called type-I and are used in almost all MOTs. Atomic MOTs have also been made using type-II transitions that have $F' \le F$~\cite{Prentiss1987, Prentiss1988, Flemming1997, Tiwari2008}. These MOTs tend to produce relatively hot clouds with low density, so have not been much used. In molecular MOTs, however, type-II transitions must be used to avoid rotational branching~\cite{Stuhl2008}. Like their atomic counterparts, these MOTs also exhibit low density and high temperature, and this has stimulated renewed interest in the cooling and trapping mechanisms at work~\cite{Tarbutt2014, Tarbutt2015, Devlin2016}, and in methods to increase the phase-space density obtained in type-II MOTs. Here, we achieve this through a conceptually simple, but counter-intuitive, change to the normal procedure---we use light that is blue-detuned from the atomic transition. We demonstrate a blue-detuned type-II MOT of $^{87}$Rb with a density exceeding $10^{11}$~cm$^{-3}$ and a temperature below 30~$\mu$K. The dimensionless phase-space density is $6 \times 10^{-6}$, about a million times higher than for a red-detuned type-II MOT~\cite{Tiwari2008}. Indeed, this phase-space density is considerably higher than usually achieved in a normal, type-I MOT, and is similar to that achieved in the best dark-SPOT MOTs~~\cite{Ketterle1993, Radwell2013}.

\begin{figure}[tb]
	\centering
	\includegraphics{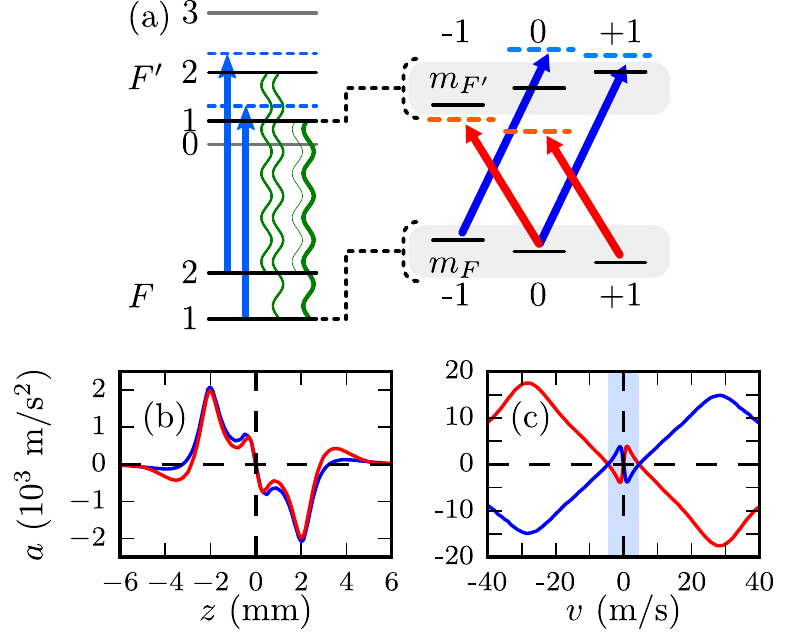}
	\caption{(a) Relevant energy levels and the transitions we drive. Relative decay rates are indicated by the thickness of the green wavy lines. The Zeeman structure of the $1\to1$ transition is shown. Trapping is obtained when the light is red-detuned and polarised to drive $\Delta m_F = -1$ transitions, or blue-detuned and polarised to drive $\Delta m_F=+1$. (b) Position-dependent acceleration curves for these two cases, showing that, near the trap centre, the force is unchanged when the polarisation and detuning are reversed. For large Zeeman splittings the symmetry is broken by the other hyperfine components. (c) Velocity-dependent acceleration curves for the two cases. For positive detuning (blue curve) there is polarisation-gradient cooling but Doppler heating, while for negative detuning (red curve) the opposite is true. (b) and (c) are calculated for a six-beam MOT using optical Bloch equations~\cite{Devlin2016} that take into account the $^{87}$Rb level structure and Zeeman shifts, and the two laser frequency components of light. The parameters used were $\delta f_{11} = \pm 11.5$~MHz, $\delta f_{22} = \pm 26$~MHz, $I_{t} = 113$~mW/cm$^{2}$ and $B'=87$~G/cm.}
	\label{RedBlueMOTCartoon}
\end{figure}

Figure \ref{RedBlueMOTCartoon} illustrates the principle of the blue-detuned MOT. The $z$-axis is defined by the magnetic field $\vec{B}$. Lasers drive the $F \rightarrow F'=1\rightarrow 1$ and $2\rightarrow 2$ D2 transitions in $^{87}$Rb. Both are type-II transitions. Provided the detunings are much smaller than the hyperfine intervals, so that each frequency component primarily drives only one transition, the position-dependent force in a MOT is unchanged when the detuning and the handedness of the light are both reversed, as illustrated in Fig.~\ref{RedBlueMOTCartoon}(b). For our level scheme, trapping is obtained when the light is red-detuned and the restoring beam, propagating towards $-z$, is polarized to drive $\Delta m_{F}=-1$ transitions, or when blue-detuned and polarized to drive $\Delta m_{F}=+1$. By contrast, the velocity-dependent force, shown in Fig.~\ref{RedBlueMOTCartoon}(c), changes sign when the detuning is reversed, irrespective of polarization choice. For type-I transitions, the Doppler and polarization-gradient components of the force have the same sign, both providing cooling for red-detuned light. For type-II transitions, these two components of the force have opposite signs, so sub-Doppler cooling requires blue-detuned light. Sub-Doppler cooling on type-II transitions has been applied to atoms~\cite{Valentin1992, Boiron1995, Hemmerich1995, Grier2013} and molecules~\cite{Truppe2017b} in optical molasses. This cooling relies on coupling between dark and bright states, induced either by motion through the changing polarization of the light field~\cite{Weidemuller1994}, or by an applied magnetic field~\cite{Shang1990, Emile1993, Gupta1994}. A recent theoretical study~\cite{Devlin2016} of these polarization gradient forces in a 3D type-II molasses shows that they are strong and can act over a wide velocity range. This can be seen in Fig.~\ref{RedBlueMOTCartoon}(c) where sub-Doppler cooling is effective for speeds below a critical velocity, $v_{\rm c}\approx 4.5$~m/s, where the force crosses zero. This is the mean speed for a thermal distribution at 80~mK. The study also showed that while polarization-gradient forces in type-I systems are suppressed by magnetic fields, for type-II systems they remain strong over the whole range of fields atoms explore in a typical MOT. Based on these findings, it was suggested~\cite{Devlin2016} that a blue-detuned type-II MOT cooled exclusively by polarization-gradient forces should be feasible, and would confine atoms just as strongly as its red-detuned counterpart while cooling to far lower temperature. Here, we demonstrate that idea.

We first load a standard red-detuned type-I $^{87}$Rb MOT from the background vapour produced by a dispenser. We form two sets of MOT beams with opposite handedness, and can switch between them using optical switches. Once the desired number of atoms has been loaded, up to a maximum of $10^{9}$, the light is switched off, the polarization handedness is reversed, the laser frequencies are stepped via an offset lock so that one frequency component is detuned by $\delta f_{11}$ from the $1\rightarrow 1$ transition and the other by $\delta f_{22}$ from $2\rightarrow 2$, and then, after a settling period of 2~ms, the light is turned back on. The axial magnetic field gradient, $B'$, is switched to a new value during this period. The six MOT beams have Gaussian intensity distributions with $1/e^{2}$ radii of 5.8~mm, producing a total peak intensity at the MOT of $I_{\rm t}$, divided approximately equally between the two frequency components. The atoms are detected by absorption imaging using light tuned close to the $2\rightarrow 3$ transition.

The fraction of atoms recaptured in the blue-detuned MOT is close to 100\% over a wide range of parameters. Nevertheless, the MOT fluorescence decreases to 20--30\% of its value in the type-I MOT. This reduction in scattering rate is expected because the sub-Doppler cooling mechanism relies on atoms spending part of their time in dark states~\cite{Weidemuller1994}. We only observe trapping when $\delta f_{11}>0$, as expected for the reversed polarization handedness. We observe a MOT for both positive and negative $\delta f_{22}$, but the temperature is higher and the density lower when  $\delta f_{22} < 0$. This shows that the $1 \rightarrow 1$ transition is mainly responsible for the MOT, which makes sense since the $F'=1$ state decays to $F=1$ with probability 5/6, whereas $F'=2$ decays to the two ground states with equal probability. We have measured the state distribution and find that, for our typical parameters, 85\%-90\% of ground-state atoms are in $F=1$. 

\begin{figure}[tb]
	\centering
	\includegraphics{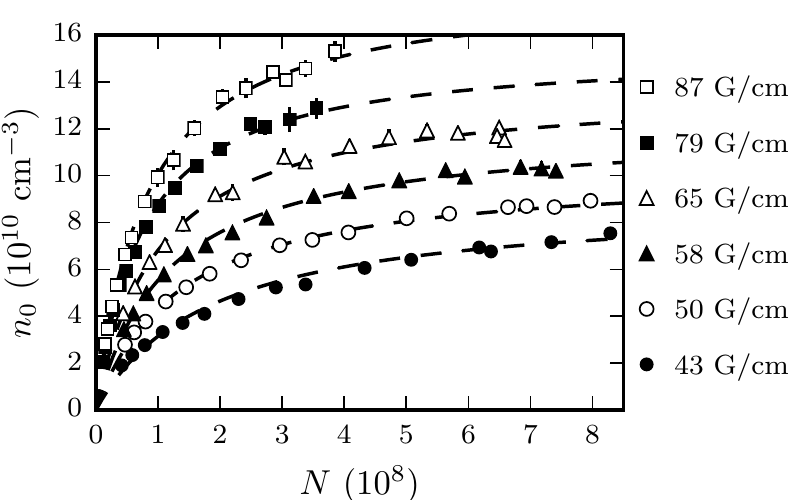}
	\caption{Peak number density, $n_0$, versus the number of atoms in the MOT, $N$, for a range of magnetic field gradients. Dashed lines are fits to $n_{0}= N/(V_{0}+N/n_{\rm max})$.
	}
	\label{Fig: nVsN}
\end{figure}

Figure \ref{Fig: nVsN} shows how the peak number density, $n_{0}$, varies with the number of trapped atoms, $N$, for various $B'$. The density initially increases linearly with $N$, and can be described by $n_{0} = N/V_{0}$ where $V_{0}$ is the volume for small $N$. For higher $N$, the density saturates towards a maximum value, which we attribute to photon re-scattering---a photon scattered by one atom can be re-scattered by another, introducing an effective repulsive force which counteracts the MOT confinement, giving a maximum attainable density, $n_{\rm max}$, independent of $N$~\cite{Walker1990, Sesko1991}. We fit the data to the model $n_{0} = N/(V_{0}+N/n_{\rm max})$, with $V_{0}$ and $n_{\rm max}$ as free parameters. This model fits well over the range of $N$ and $B'$ explored. We find a linear relationship between $n_{\rm max}$ and $B'$, as expected~\cite{Petrich1994}, with a gradient of $2.1(1)\times10^9$~cm$^{-3}$/(G/cm). Provided the MOT temperature is independent of $B'$, which it is over this range (see below), and the restoring force is linear in the displacement with a slope proportional to $B'$, we expect the relationship $V_{0}\propto (B')^{-3/2}$. We measure $V_{0}\propto (B')^{-p}$ with $p=2.2(1)$. For a standard type-I Rb MOT, operated with $B'\approx 10$~G/cm, the density limit is about $2 \times 10^{10}$~cm$^{-3}$. Using a compressed MOT (CMOT), this can be increased to $5 \times 10^{11}$~cm$^{-3}$ by ramping up $B'$ to values similar to those we use here~\cite{Petrich1994}. Remarkably, despite its small spring constant (see below), the blue-detuned MOT has $n_{\rm max}$ similar to that of a type-I CMOT. We attribute this to reduced photon re-scattering due to the lower scattering rate.

\begin{figure}[tb]
	\centering
	\includegraphics[trim ={3mm 0mm 0mm 0mm}, clip]{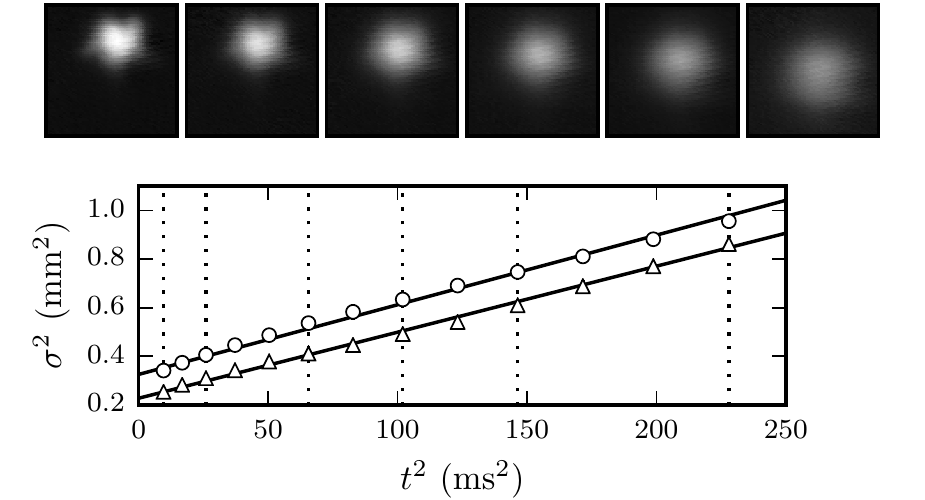}
	\caption{Ballistic expansion of atoms released from the blue-detuned MOT. The parameters are $\delta f_{11} = 35$~MHz, $\delta f_{22} = 11.5$~MHz and $B' = 87$~G/cm. The MOT is loaded at $I_{t}=130$~mW/cm$^{2}$ and $I_{t}$ is then reduced to 22~mW/cm$^{2}$ for 5~ms before the atoms are released. Images are 5~mm$~\times~$5~mm and correspond to the times indicated by the dotted lines in the lower panel. The lower panel shows the rms widths in the radial (circles) and axial (triangles) directions. The fitted temperatures are $T_{\rho} = 30(1)$~$\mu$K and $T_{z} = 29 (1)$~$\mu$K.
	}
	\label{Fig: BallisticExpansion}
\end{figure}

We turn next to the temperature. Figure \ref{Fig: BallisticExpansion} shows typical images of an expanding cloud of $N=2.4(1) \times 10^{8}$ atoms released from the blue-detuned MOT. The axial and radial density distributions of these clouds fit well to a Gaussian model. The initial rms widths in these directions are $\sigma_{z} = 0.43(2)$~mm and $\sigma_{\rho} = 0.49(2)$~mm, giving a peak density of $n_{0} = N/[(2\pi)^{3/2}\sigma_{\rho}^{2}\sigma_{z}]=1.5 \times 10^{11}$~cm$^{-3}$. The geometric mean temperature found from the expansion is $T=30(1)$~$\mu$K. This is lower than typically obtained in a normal Rb MOT, especially for such a high density. In density-limited red-detuned MOTs, atoms are heated by photon re-scattering~\cite{Hillenbrand1994} and are pushed out by internal radiation pressure to regions of larger magnetic field where sub-Doppler cooling is inhibited. In the blue-detuned MOT, the reduced scattering rate reduces this heating, and sub-Doppler cooling is more robust to magnetic fields~\cite{Devlin2016}, so lower temperatures can be reached. 

\begin{figure}[tb]
	\centering
	\includegraphics{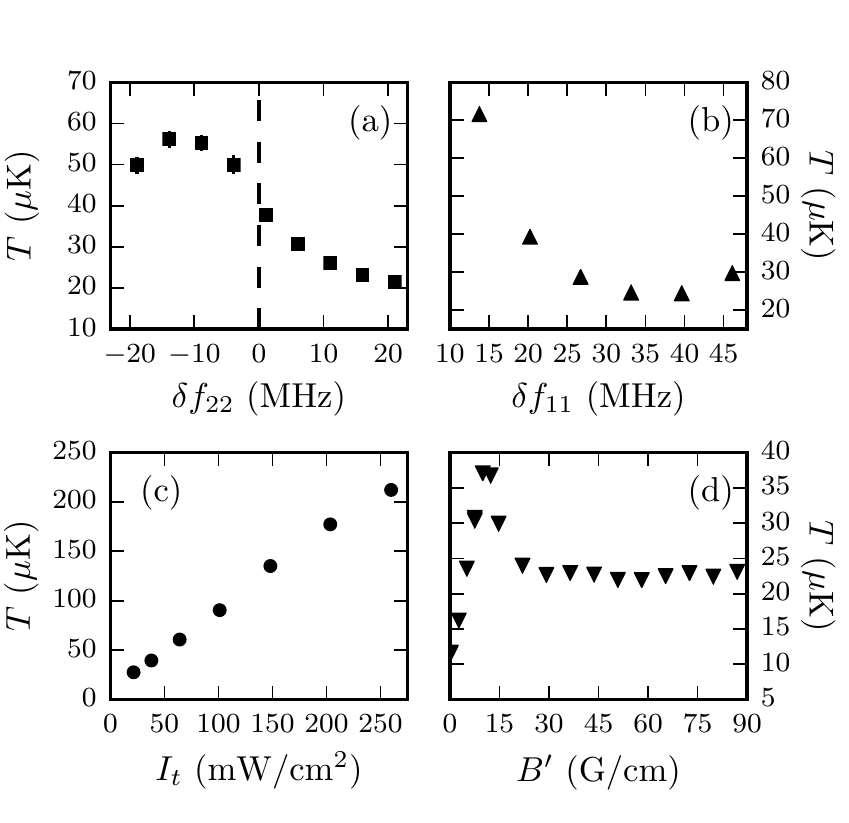}
	\caption{(a) Temperature versus $\delta f_{22}$ when $\delta f_{11} = 35$~MHz, $I_{t} = 21$~mW/cm$^{2}$ and $B'=40$~G/cm. (b) Temperature versus $\delta f_{11}$ when $\delta f_{22} = 11.5$~MHz, $I_{t} = 19$~mW/cm$^{2}$ and $B'=40$~G/cm. (c) Temperature versus $I_{t}$ when $\delta f_{11} = 35$~MHz,  $\delta f_{22} = 11.5$~MHz and $B'=88$~G/cm. After loading the MOT at full intensity, $I_{t}$ is held at the reduced value for 5~ms before measuring the temperature. (d) Temperature versus $B'$ when $\delta f_{11} = 35$~MHz,  $\delta f_{22} = 11.5$~MHz and $I_{t} = 28$~mW/cm$^{2}$.}
	\label{Fig: TemperatureProperties}
\end{figure}

Figure \ref{Fig: TemperatureProperties} shows how the temperature depends on the key parameters. Figure \ref{Fig: TemperatureProperties}(a) shows that the temperature is low for both positive and negative $\delta f_{22}$ but that the lowest temperatures are found when $\delta f_{22} > 0$, as expected. Figure \ref{Fig: TemperatureProperties}(b) shows that the temperature decreases sharply as $\delta f_{11}$ increases beyond $10$~MHz, reaching a minimum near $35$~MHz. We expect the temperature to be inversely proportional to the damping coefficient and proportional to the momentum diffusion constant, which is in turn proportional to the scattering rate. The damping coefficient is predicted to be intensity-independent~\cite{Devlin2016} while the scattering rate increases with intensity, so we expect $T$ to increase with $I_{t}$. Figure \ref{Fig: TemperatureProperties}(c) shows that the relation is a linear one over the range of $I_{t}$ explored.  Figure \ref{Fig: TemperatureProperties}(d) shows that the temperature initially increases with $B'$, but then gradually declines again beyond 15~G/cm, settling to a constant value for $B'>30$~G/cm. This behaviour reflects the velocity-dependent force curves for various magnetic fields given in Ref.~\cite{Devlin2016}. At low $B$, sub-Doppler cooling relies on motion-induced non-adiabatic transitions between dark and bright states~\cite{Weidemuller1994}, while at higher $B$ Larmor precession between bright and dark states can take over~\cite{Emile1993, Gupta1994}. 

The position- and velocity-dependent forces are quantified in terms of the spring constant, $\kappa$, and damping coefficient, $\alpha$: $m\ddot{z} = - \kappa z - \alpha \dot{z}$. We obtain $\kappa$ from the equipartition theorem which gives $\kappa_{\rho(z)}\sigma_{\rho(z)}^{2} = k_{B}T$. This expression is valid for a harmonic trap in the limit of low $N$ where multiple photon scattering can be neglected.  We use the values of $V_{0}$ from the fits shown in Fig.~\ref{Fig: nVsN} to determine $\sigma_\rho^2\sigma_z$, and use the relation $\sigma_{\rho} = \sqrt{2}\sigma_{z}$ to obtain $\sigma_{\rho}$. Provided the MOT beams are carefully aligned and their intensities well balanced, we find this relation to be accurate for low $N$. In this same limit, we measure equal axial and radial temperatures and find that a ten-fold increase in $N$ changes the temperature by less than 10\%. From measurements of $V_{0}$ and $T$ we obtain $\kappa_{z}=5(2)\times10^{-20}$~N/m when $I_{t} = 113$~mW/cm$^{2}$, $B' = 87$~G/cm, $\delta f_{11} = 35$~MHz and $\delta f_{22}=11.5$~MHz. The large uncertainty reflects the range of $\kappa_{z}$ measured for various beam alignments and intensity imbalances, all giving small, roughly spherical MOTs. From a theoretical curve similar to Fig.~\ref{RedBlueMOTCartoon}(b), calculated for the exact parameters used here, we predict $\kappa_{z}=15(6)\times10^{-20}$~N/m, roughly 3 times larger than measured. The measured value is 50 times smaller than found for a normal Rb MOT operated at $B'\approx 20$~G/cm~\cite{Wallace1994}. From data reported for a red-detuned type-II MOT of $^{85}$Rb~\cite{Tiwari2008}, we infer $\kappa_{z}=8\times 10^{-20}$~N/m, consistent with our value. Our value is also similar to those found for dc and rf MOTs of SrF and CaF molecules~\cite{McCarron2015, Norrgard2016, Williams2017, Anderegg2017}. With $\kappa_z$ fixed, we obtain $\alpha$ from the over-damped relaxation of the MOT following a rapid displacement of the trap centre.  For the same parameters as above, we find $\alpha/m=4.2(6)\times10^3$~s$^{-1}$. For these parameters, the theoretical curve similar to Fig.~\ref{RedBlueMOTCartoon}(c) gives $\alpha/m=3.9(2)\times10^3$~s$^{-1}$, consistent with the measurement.

\begin{figure}[t]
	\centering
	\includegraphics[trim ={0mm 0mm 0mm 5mm}, clip]{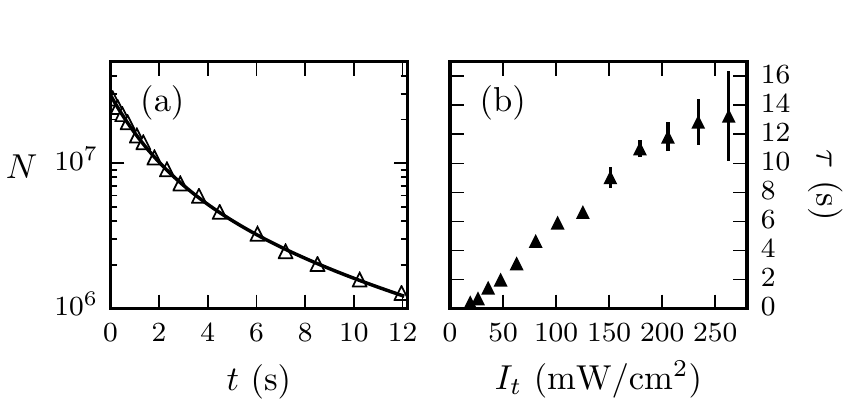}
	\caption{(a) Number of atoms in the MOT as a function of time. The blue-detuned light is turned on at $t=0$. Parameters are $I_{t}=240$~mW/cm$^{2}$, $\delta f_{11} = 26$~MHz, $\delta f_{22} = 12$~MHz, $B'=48$~G/cm. Line: fit to the model described in the text. (b) Lifetime, $\tau$, measured for small $N$, as a function of $I_{t}$. Parameters are $\delta f_{11} = 26$~MHz, $\delta f_{22} = 12$~MHz, $B'=39$~G/cm.}
	\label{Fig:Lifetime}
\end{figure}

Figure \ref{Fig:Lifetime}(a) shows how the number of atoms decays over time. We describe the atom loss by
\begin{equation}
\frac{dN}{dt}=-\gamma N - \beta \int n^{2}d^{3}r.
\end{equation}
The first term on the right describes loss due to background gas collisions, while the second describes loss due to collisions between atoms in the MOT. Using a Gaussian density distribution with peak density $n_{0}=N/(V_{0}+N/n_{\rm max})$ we obtain
\begin{equation}
\frac{dN}{dt}=-\gamma N - \frac{\beta}{2^{3/2}} \frac{N^{2}}{V_{0}+N/n_{\rm max}}.
\end{equation}
We fit the data to the solution of this equation, with $V_{0}$ and $n_{\rm max}$ fixed to the values found from the fits in Fig.~\ref{Fig: nVsN}. The data fit well to this model, giving $\gamma = 0.075(3)$~s$^{-1}$ and $\beta = 1.75(3)(35)\times10^{-10}$~cm$^{3}$~s$^{-1}$. For $\beta$, the first bracketed number is the statistical uncertainty in the measurement, while the second is the systematic uncertainty due to the uncertainty in $N$. In the limit of small $N$, the trap lifetime is $\tau = 1/\gamma =13.3(5)$~s. This is about 4 times longer than the characteristic loading time of the type-I MOT for the same $I_{t}$, implying that the lifetime of the blue-detuned MOT is longer than the type-I MOT at this $I_{t}$. This is consistent with our observation that the excited-state fraction in the blue-detuned MOT is about a quarter that of the normal MOT, together with the previous observation that collisions with background Rb eject excited-state atoms at 3 times the rate of ground-state atoms~\cite{Anderson1994}.  Figure \ref{Fig:Lifetime}(b) shows how the lifetime depends on $I_{t}$ in the low-$N$ limit. The observed increase of $\tau$ with $I_{t}$ is surprising, since the trap depth is far smaller than the temperature of the background gas, but may be due to glancing collisions which are effective at ejecting atoms at low $I_{t}$, where the trap depth is low, but less so as $I_{t}$ increases. A similar observation has been noted for a normal MOT operated at low intensity~\cite{Wallace1992}. The coefficient $\beta$ has been studied previously in $^{87}$Rb MOTs~\cite{Wallace1992, Gensemer1997}. Values similar to ours were measured at low intensity, where the trap depth was lower than the energy released in a hyperfine-changing collision. With increasing intensity $\beta$ was observed to fall as the trap depth increased, but then increased again as collisions between ground and excited atoms started to dominate. For intensities exceeding 15~mW/cm$^{2}$, a roughly constant $\beta \approx 2 \times 10^{-12}$~cm$^{3}$~s$^{-1}$ was measured, 100 times smaller than our value. We suggest that hyperfine-changing collisions dominate in our MOT for all $I_{t}$, because of its low trap depth, and that this is the reason for the large $\beta$. We note that for a red-detuned type-II MOT of $^{85}$Rb, $\beta = 9.1(7) \times 10^{-9}$~cm$^{3}$~s$^{-1}$ has been reported~\cite{Tiwari2008}.

The MOT shown in Fig.~\ref{Fig: BallisticExpansion} has a phase-space density of $6 \times 10^{-6}$, the maximum we have observed in these experiments. This is a million times higher than previously reported for a red-detuned type-II MOT of Rb~\cite{Tiwari2008}. For molecular MOTs, which always use type-II transitions, a switch from red to blue detuning offers a similarly great leap in phase-space density. The blue-detuned MOT also compares favourably to type-I alkali MOTs where the highest phase-space densities are obtained using the dark SPOT technique~\cite{Ketterle1993}. Our phase-space density is about the same as the highest achieved by that method~\cite{Radwell2013}. We note that even higher phase-space density was recently achieved for Sr atoms by using several stages of laser cooling~\cite{Bennetts2017}. Blue-detuned MOTs seem advantageous for efficient loading of atoms and molecules into conservative traps, including optical tweezer traps and chip-scale traps, where dissipation, low temperature, and confinement to a small volume are all needed. Our MOT offers new opportunities to study the control of ultracold collisions, especially the suppression of inelastic processes by optical shielding~\cite{Bali1994, SanchezVillicana1995, Suominen1996} which should be effective for near-resonant blue-detuned light. It can also serve as a good starting point for evaporation of atoms and molecules to quantum degeneracy.

We are grateful for helpful discussions with Ed Hinds and Jeremy Hutson. This research has received funding from EPSRC under grants EP/M027716/1 and EP/P01058X/1.
\bibliography{BlueMOTBib}
\end{document}